\documentstyle[prl,aps,multicol,amssymb,epsfig]{revtex}
\title{Dissipation-assisted quantum gates with cold trapped ions}
\author{Almut Beige}
\address{Max-Planck-Institut f\"ur Quantenoptik, Hans-Kopfermann-Str.~1,  
85748 Garching, Germany.}
\date{\today}

\begin{document}

\maketitle  
\draft

\begin{abstract}
\begin{center}
\parbox{14cm}
{It is shown that a two-qubit phase gate and SWAP operation between ground states 
of cold trapped ions can be realised in one step by simultaneously applying two 
laser fields. Cooling during gate operations is possible without perturbing the 
computation and the scheme does not require a second ion species for sympathetic 
cooling. On the contrary, the cooling lasers even stabilise the desired time evolution 
of the system. This affords gate operation times of nearly the same order of magnitude 
as the inverse coupling constant of the ions to a common vibrational mode.}
\end{center}
\end{abstract}

\vspace*{0.2cm}
\noindent

\begin{multicols}{2}
For many years now ion trapping technology has been one of the standard techniques 
for investigating quantum phenomena with single particles. 
Paul traps or more complex structures have been used to create chains of several 
hundred ions\cite{drewsen}. Many schemes for 
implementing quantum gates using trapped ions have been proposed. Some of them 
require cooling into the vibrational ground state\cite{cirac,mon,jonathan,sasura}. 
The feasibility of this has already been experimentally demonstrated\cite{wine}. 
Other schemes can even be implemented with ``hot'' ions\cite{poy} and have been 
applied to entangle up to four ions\cite{sackett} and to observe violation of 
Bell's inequality with single atoms\cite{david}.

This paper proposes an alternative scheme for realising a phase gate 
and SWAP operation with cold trapped ions in one step 
by simultaneously applying two laser fields. Compared with other 
schemes\cite{cirac,jonathan}, the experimental effort for quantum computing can 
thus be greatly reduced. Ideally, the system remains during the whole 
gate operation in the ground state of a common vibrational mode. Thus cooling 
of this mode does not perturb the computation. On the contrary, it is 
shown that cooling can even improve the gate fidelity significantly. This affords 
gate durations of nearly the same order of magnitude as the inverse coupling 
constant of the ions to the common vibrational mode and there is no need to include 
a different ion species for sympathetic cooling\cite{morigi} in the trap. 
In addition, the scheme only requires good control over one Rabi frequency.
The size of the coupling strength of the ions to the vibrational mode does not enter 
the effective time evolution of the qubits.

Each qubit is obtained from the atomic ground states $|0 \rangle$ and $|1 \rangle$ of 
one ion, while the common vibrational mode is cooled down to its ground state 
$|0_{\rm vib}\rangle$. To establish coupling between qubits, we use a metastable state 
$|2 \rangle$ and a strong laser field with Rabi frequency $\Omega_2$ 
detuned by the phonon frequency $\nu$, as 
shown in Figure \ref{fig1}. The phase gate that adds a minus sign 
to the amplitude of the qubit state $|01 \rangle$ then requires in addition only 
a weak laser pulse with the Rabi frequency $\Omega$ coupling resonantly to the 
0-2 transition of ion 1. Let us denote the creation operator of a single phonon 
in the mode as $b^\dagger$ and introduce the coupling constant of the ions to the 
common vibrational mode as $g_2 = {1 \over 2} \eta \Omega_2$. (Here $\eta$ 
is the Lamb-Dicke parameter characterising the steepness of the trap.)
The Hamiltonian of the system within the dipole and the rotating wave 
approximation and in the interaction picture with respect to the free Hamiltonian 
is~then~given~by
\begin{eqnarray} \label{hlaserI}
H &=& \sum_{i=1}^2 {\rm i} \hbar g_2 \, |1 \rangle_i\langle 2| \, b^\dagger 
+ {\textstyle {1 \over 2}} \hbar \Omega \, |0 \rangle_1 \langle 2| + {\rm h.c.}
\end{eqnarray}
Here the Lamb-Dicke regime and the condition $g_2^2 \ll \nu^2$ have been assumed, as 
in\cite{cirac}. 

\vspace*{-0.2cm}
\noindent\begin{minipage}[ht]{3.38truein}
\begin{center}
\begin{figure}
\epsfig{file=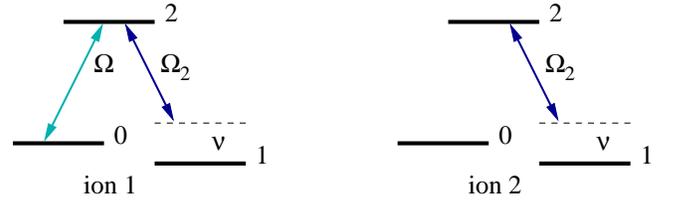,width=3.38truein} \\[0.2cm]
\caption{Level scheme for implementation of a phase gate. A strong laser field 
with detuning $\nu$ establishes coupling between the ions via a common vibrational mode. 
In addition, a laser pulse individually addressing the 0-2 transition of ion 1 is 
required.} \label{fig1}
\end{figure}
\end{center}
\end{minipage}
\vspace*{0.2cm}

Let us first assume that the coupling constant $g_2$ is a few orders of 
magnitude larger than the Rabi frequency $\Omega$. Then there are two different time 
scales in the system and the time evolution can be calculated to very good 
approximation by adiabatic elimination. To do so, the amplitude of the state with 
$n$ phonons in the vibrational mode and the ions in $|ij \rangle$ is denoted as  
$c_{nij}$. Only the coefficients of the qubit states, $c_{000}$, $c_{001}$, $c_{010}$ and 
$c_{011}$, and of the entangled state $|0a\rangle$ with $|a\rangle \equiv ( 
|12 \rangle - |21\rangle )/\sqrt{2}$ change slowly in time. Their time evolution 
is given by 
\begin{eqnarray} \label{cdgls}
\dot{c}_{000} &=& - {\textstyle {{\rm i} \over 2}} \, \Omega \, c_{020}~, \nonumber \\
\dot{c}_{001} &=& {\textstyle {{\rm i} \over 2\sqrt{2} }} \, \Omega \, 
\big( c_{0a} - c_{0s} \big) ~, \nonumber \\
\dot{c}_{0a} &=& {\textstyle {{\rm i} \over 2\sqrt{2} }} \, \Omega \, c_{001} \nonumber \\
\dot{c}_{010} = \dot{c}_{011} &=& 0 ~.
\end{eqnarray}
Setting the derivatives of all other coefficients equal to zero yields
$c_{020}=c_{0s}=0$. For $\Omega \ll g_2$, the time evolution of the system (\ref{cdgls})
can thus be summarised in the effective Hamiltonian
\begin{eqnarray} \label{eff}
H_{\rm eff} &=& - {\textstyle {1 \over 2 \sqrt{2}}} \,  \hbar \Omega \, \big[ \, 
|001 \rangle \langle 0a| + {\rm h.c.} \, \big] ~.
\end{eqnarray}
If the duration of the laser pulse equals $T = 2\sqrt{2} \pi/\Omega$,
then the resulting evolution is the desired phase gate. As the effective 
Hamiltonian $H_{\rm eff}$ and the gate operation time $T$ are independent of the 
coupling constant $g_2$, the proposed scheme is widely protected against 
fluctuations of this system parameter.

Deviations from the time evolution (\ref{eff}) arise because population accumulates 
unintentionally in the states $|110\rangle$ and $|111\rangle$. In first order in 
$\Omega/g_2$ one has
\begin{equation} \label{fast}
c_{110} = - {{\rm i}\Omega \over 2 g_2} \, c_{000} ~,~
c_{111} = - {{\rm i}\Omega \over 4 g_2} \, c_{001}
\end{equation}
and the fidelity of the proposed phase gate equals 
\begin{eqnarray} \label{F}
F(T,|\psi \rangle) &=& 1 - {\Omega^2 \over 4 g_2^2} \, \big[ \, |c_{000}|^2
+ {\textstyle {1 \over 4}} \, |c_{001}|^2 \, \big] ~.
\end{eqnarray} 
Figure \ref{fig2} compares this fidelity with the exact solution resulting 
from numerical integration of the time evolution (\ref{hlaserI}). Good agreement is 
only found for $\Omega \ll g_2$. For somewhat larger Rabi frequencies, in general 
fidelities worse than the result predicted by (\ref{F}) are obtained due to 
non-adiabaticity. Thus the 
fidelity is above $99\,\%$ for all initial states if $\Omega < 0.1 \, g_2$ is 
chosen. This corresponds to gate operation times $T > 90/g_2$. In the following we 
aim at enlarging the parameter regime for which the fidelity is at least $99\,\%$
\cite{note}.

\vspace*{-0.2cm}
\noindent\begin{minipage}{3.38truein}
\begin{center}
\begin{figure}
\epsfig{file=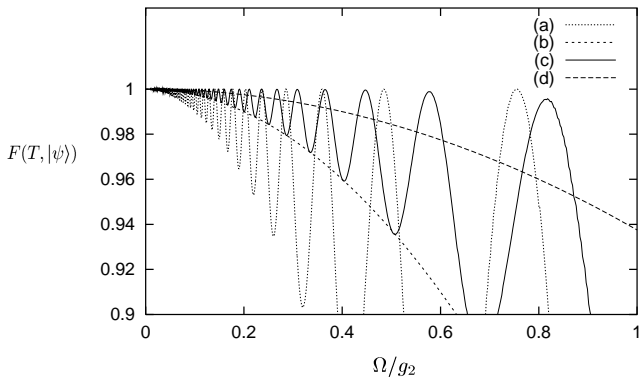,width=3.38truein} \\[0.2cm]
\caption{The fidelity of a single phase gate as a function of $\Omega/g_2$ for the 
initial qubit states $|00\rangle$ (a,b) and $|01\rangle$ (c,d). The curves (a,c) 
result from an exact solution of the time evolution (\ref{hlaserI}) while the curves
(b,d) result from (\ref{F}). For the states $|10\rangle$ and  $|11\rangle$ optimal 
fidelities $F\equiv 1$ are obtained.} \label{fig2}
\end{figure}
\end{center}
\end{minipage}
\vspace*{0.2cm}

Dominating error source in the scheme is heating. To avoid this, the gate should 
be performed fast. Therefore we assume in the following that $\Omega$ is of nearly 
the same size as the coupling constant $g_2$. As can be seen from above, increasing 
$\Omega$ leads to the population of unwanted states. To reduce the error rate of 
the scheme one could therefore measure the population of states with $n>0$ 
at the end of each gate. Under the condition that 
no population is found in these states, the system gets projected back onto the 
subspace with $n=0$ and the fidelity of the prepared state increases. 
In case of the detection of an error, the gate failed and the whole computation has 
to be repeated. 

To realise such an error detection measurement one could use a laser field that 
couples the atomic ground state $|1\rangle$ 
with detuning $\nu$ to an auxiliary state $|3\rangle$ and another laser that 
excites the atomic 1-2 transition with the same detuning. To assure that populating 
level 3 leads to the emission of many photons, an even stronger laser should 
couple $|3 \rangle$ to a rapidly decaying fourth level which decays into 
$|3\rangle$ with a high spontaneous decay rate. Gate failure 
leads thus to an effect which is known as a ``macroscopic light period''\cite{dehmelt}, 
i.e.~fast Rabi oscillations between level 3 and 4 accompanied by photon emission at 
a high rate (for details see\cite{neu}). This can easily be detected and 
the whole computation can be repeated, if necessary. 

\vspace*{-0.2cm}
\noindent\begin{minipage}[ht]{3.38truein}
\begin{center}
\begin{figure}
\epsfig{file=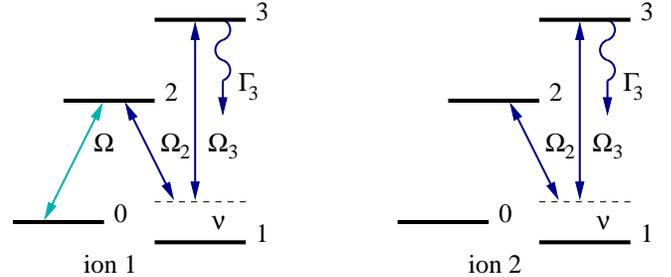,width=3.38truein} \\[0.2cm]
\caption{Level scheme of a dissipation-assisted phase gate with cold trapped ions. 
In addition to the basic setup shown in Figure \ref{fig1}, a strong laser field couples 
the state $|1\rangle$ with detuning $\nu$ to an auxiliary atomic state $|3\rangle$. 
Populating this level leads to the emission of photons with a rate $\Gamma_3$.} \label{fig3}
\end{figure}
\end{center}
\end{minipage}
\vspace*{0.2cm}

However, such an error detection 
measurement would take much longer than the inverse of the coupling constants 
of the ions to the vibrational mode and much longer than $T$. Hence, error 
detection does not help to decrease the gate operation time for a given 
minimum fidelity $F$. To shorten the gate duration without reducing the fidelity 
more than predicted by (\ref{F}), we propose that 
the desired time evolution (\ref{eff}) be stabilised during the gate performance 
by using dissipation. This is achieved by continuously applying the laser fields, 
proposed for the implementation of error detection, as shown in Figure \ref{fig3}. 
For simplicity, the strong laser coupling to 
the fourth level and spontaneous decay from this level have been combined into a 
single decay rate assigned to the metastable state $|3 \rangle$. Let us denote 
this spontaneous decay rate as $\Gamma_3$, while 
$g_3 = {1 \over 2} \eta \Omega_3$ is the coupling strength 
of the atomic 1-3 transition to the vibrational mode. In the following, 
$g_3 \sim g_2$ and $10 \, g_2 < \Gamma_3 < 100 \, g_2$ is assumed.

More general, any process that indicates whether the phonon mode 
is excited or not can serve as an error detection measurement. This applies to 
ground state cooling\cite{ground} because populating the vibrational mode leads 
to the emission of photons at a high rate while no emission takes place if the 
ions are in the vibrational ground state. Thus level 3 and the additional strong 
laser field in Figure \ref{fig3} could as well be replaced by the cooling laser 
setup. Indeed, continuous cooling can 
improve the fidelity of the performed gate operation. However, for simplicity, 
the continuous read out of the phonon mode is in the following modeled as 
shown in Figure \ref{fig3}.

Basic mechanism of the improved scheme is that observing for emitted 
photons implements a (conditional) no-photon time evolution, thus resulting in 
continuous damping of the population in unwanted states. As long as the amplitudes 
$c_{020}$ and $c_{0s}$ are negligible, the time evolution of the other states 
resembles the desired phase gate, as can be seen from (\ref{cdgls}). In addition,
we show that the population that now accumulates in the states $|110\rangle$ 
and $|111 \rangle$ is about the same as predicted for 
an adiabatic process and the fidelity of the 
phase gate coincides with $F(T,|\psi \rangle)$ given in (\ref{F}) to a very good 
approximation. The price one has to pay for this improvement of the precision 
of the gate is that photon emission might occur with a small probability. Then the gate 
operation would have failed.

To describe the time evolution of the system under the condition of no photon emission, 
we use in the following the Schr\"odinger equation with the conditional Hamiltonian 
$H_{\rm cond}$. As predicted by the quantum jump approach\cite{HeWi11}, the norm of a 
vector developing with this non-Hermitian Hamiltonian decreases in general with time 
and 
\begin{equation} \label{pipi}
P_0(T,|\psi \rangle) = \| \, U_{\rm cond} (T,0) \, |\psi \rangle \, \|^2
\end{equation}
is the probability of no photon emission in $(0,T)$ if $|\psi\rangle$ is the initial 
state of the system. For the level configuration shown in Figure \ref{fig3} the 
conditional Hamiltonian equals
\begin{eqnarray} \label{hcond}
H_{\rm cond} &=& \sum_{i=1}^2 \sum_{j=2}^3 {\rm i} \hbar g_j \, 
|1 \rangle_i\langle j| \, b^\dagger + {\textstyle {1 \over 2}} \hbar \Omega \, 
|0 \rangle_1 \langle 2| + {\rm h.c.} \nonumber \\
&& - {\textstyle{{\rm i} \over 2}} \hbar \Gamma_3 \, |3 \rangle_i\langle 3| ~. 
\end{eqnarray}
Because of the different time scales of the scheme, the no-photon time evolution of the 
system can again be calculated by adiabatic elimination. This yields the same effective 
Hamiltonian as in (\ref{eff}). In first order in $\Omega/g_2$, population accumulates 
unintentionally in the states $|020\rangle$, $|110\rangle$, $|030\rangle$, 
$|0s \rangle$ $|111\rangle$, $|013\rangle$ and $|031\rangle$ and it is
\begin{eqnarray} \label{fast2}
\big( c_{0s},c_{111},c_{013},c_{031} \big) &=&   
- {{\rm i} \Omega \over 2 g_2} \, \Big( {\sqrt{2} g_3^2 \over g_2 \Gamma_3}, 
{1 \over 2}, {g_3 \over \Gamma_3}, {g_3 \over \Gamma_3} \Big) \, c_{001} ~,  \nonumber \\ 
\big( c_{020},c_{110},c_{030} \big) &=&   
- {{\rm i} \Omega \over g_2} \, \Big( {g_3^2 \over g_2 \Gamma_3}, {1 \over 2}, 
- {g_3 \over \Gamma_3} \Big) \, c_{000} ~.
\end{eqnarray}
To optimise the fidelity, $\Gamma_3$ should be much larger than $g_3$ so that all 
coefficients proportional $g_3/\Gamma_3$ become negligible. In this case,  
$F(T,|\psi\rangle)$ becomes the fidelity calculated in (\ref{F}) to a very good 
approximation.

That this is indeed the case is shown in Figure \ref{fig4} which results from a 
numerical solution of the time evolution (\ref{hcond}). As expected, the 
dissipation channel continuously introduced in the system 
stabilises the desired time evolution and corrects for errors resulting from the 
non-adiabaticity of the scheme if $\Omega$ becomes of about the same order of 
magnitude as $g_2$. The gate fidelity (\ref{F}) applies now to a much wider 
parameter regime. Fidelities above $99\,\%$ are obtained if 
$\Omega < 0.18 \, g_2$ is chosen. However, for too big spontaneous decay rates 
$\Gamma_3$, the damping of unwanted amplitudes becomes ineffective which is why 
$\Gamma_3 < 100 \, g_2$ has been assumed.

\vspace*{-0.2cm}
\noindent\begin{minipage}{3.38truein}
\begin{center}
\begin{figure}
\epsfig{file=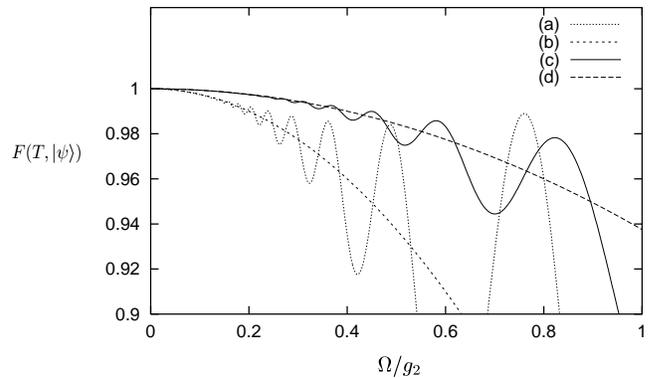,width=3.38truein} \\[0.2cm]
\caption{Fidelity of a single phase gate under the condition of no photon emission 
as a function of $\Omega/g_2$ for the initial qubit states $|\psi\rangle = |00\rangle$
(a,b) and $|\psi\rangle = |01\rangle$ (c,d). Deviations from the time evolution
with $g_3 = \Gamma_3 =0$ are continuously damped away with $g_3=g_2$ and 
$\Gamma_3=20\,g_2$ (a,c) and the fidelity is close to the theoretically predicted 
fidelity (\ref{F}) assuming adiabaticity (b,d).} \label{fig4}
\end{figure}
\end{center}
\end{minipage}
\vspace*{0.2cm}

Using the coefficients $c_{0s}$ and $c_{020}$ given in (\ref{fast2}) and the 
differential equations that govern the time evolution of the qubit states and 
the entangled state $|0a\rangle$ (the same as (\ref{cdgls})), the unnormalised 
state of the system at the end of the gate operation under the condition of no 
photon emission can be calculated up to first order in $\Omega/g_2$. 
Its norm squared equals the gate success rate (\ref{pipi}) and 
\begin{eqnarray} \label{P0}
P_0(T,|\psi \rangle) &=& 1 - 2 \sqrt{2} \pi \, {\Omega g_3^2 \over g_2^2 \Gamma_3} \, 
\big[ \, |c_{000}|^2 + {\textstyle {1 \over 4}} \, |c_{001}|^2 \, \big] ~.
\end{eqnarray} 
This is in good agreement with the numerical results shown in Figure \ref{fig5}.
For example, for $\Omega < 0.1 \, g_2$ and $\Gamma_3 = 20 \, g_2$ one has 
$P_0(T,\psi) > 95 \, \%$, independent of the initial state of the system.
The probability for photon emission during the gate operation is of the order of 
$\Omega/g_2$ and for $\Omega g_3^2 /(g_2^2 \Gamma_3) \ll 1$ close to unity.
Gate failure might be a bit more likely than for quantum error detection
(see Figure \ref{fig5}). 
However, no additional time is required which would increase the sensitivity of the 
scheme with respect to heating.

\vspace*{-0.2cm}
\noindent\begin{minipage}{3.38truein}
\begin{center}
\begin{figure}
\epsfig{file=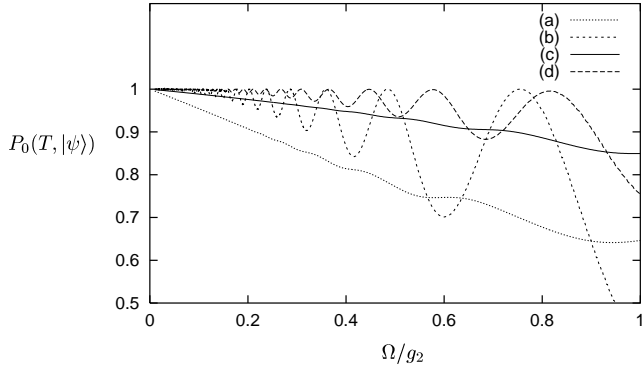,width=3.38truein} \\[0.2cm]
\caption{Success rate of a single phase gate as a function of $\Omega/g_2$ for 
the initial qubit states $|\psi\rangle = |00\rangle$ (a,b) and 
$|\psi\rangle = |01\rangle$ (c,d) for the case of continuous monitoring of the system 
(a,c) and the case of only a single error detection measurement at the end of the 
gate (b,d).} 
\label{fig5}
\end{figure}
\end{center}
\end{minipage}
\vspace*{0.2cm}

Another two-qubit gate that can easily be realised with the same experimental setup 
is SWAP operation. This gate exchanges the states of two qubits without the 
corresponding ions having to change their places physically. 
Compared with the above phase gate, its implementation does 
not require individual laser addressing. To realise SWAP operation, the laser field 
with Rabi frequency $\Omega$ should address both ions simultaneously. To improve the 
gate fidelity, the same ideas as described above can be used. By analogy with 
(\ref{hcond}), the conditional Hamiltonian of the system is now given by
\begin{eqnarray} \label{hcond3}
H_{\rm cond} &=& \sum_{i=1}^2 \sum_{j=2}^3 {\rm i} \hbar g_j \, 
|1 \rangle_i\langle j| \, b^\dagger + {\textstyle {1 \over 2}} \hbar \Omega \, 
|0 \rangle_i \langle 2|  + {\rm h.c.} \nonumber \\
&& - {\textstyle{{\rm i} \over 2}} \hbar \Gamma_3 \, |3 \rangle_i\langle 3| ~.
\end{eqnarray}
Proceeding as above leads to the effective Hamiltonian 
\begin{equation}
H_{\rm eff} = {\textstyle {1 \over 2 \sqrt{2}}} \, \hbar \Omega \,
\big[ \, - |001 \rangle \langle 0a| + |010 \rangle \langle 0a| + {\rm h.c.} \, \big]
\end{equation}
and if $T = 2 \pi/\Omega$, then
the time evolution of the system exchanges the amplitudes of the qubit states 
$|01\rangle$ and $|10\rangle$ of the initial state. Deviations of the fidelity from unity 
are, for the same parameter regime as considered above, about the same size as for 
the proposed phase gate since implementation of the two gates is very similar. 
The same applies to the improvement of the gate precision that can be achieved with 
the help of dissipation and to the gate success rates.

Summarising, we have shown that dissipation can be used to construct fast, simple 
and precise gates for quantum computing. As an example we discussed the 
implementation of a two-qubit phase gate and SWAP operation with cold trapped ions.
Here the parameter regime has been chosen such that auxiliary decay channels,
resulting from continuous cooling of the ions, stabilise the desired adiabatic 
time evolution of the system and improve the fidelity of the gate operation
significantly. The more general regime where strong ground state cooling  
allows the implementation of arbitrary single laser pulse gates will be 
discussed elsewhere\cite{neu}. \\

{\em Acknowledgment.} The author would like to thank W. Lange, C. Marr, J. Pachos 
and H. Walther for interesting discussions and for careful reading of the manuscript.

\narrowtext

\end{multicols}

\begin{references}
\bibitem{note}
For specific values of $\Omega$, for example $\Omega = 0.8 \, g_2$, fidelities above 
$96 \,\%$ are achieved even if $\Omega$ is about the same size as $g_2$.
\end{references}
\end{document}